\documentclass[fleqn,usenatbib]{mnras}

\usepackage{newtxtext,newtxmath}
\usepackage[T1]{fontenc}

\usepackage{graphicx}	% Including figure files
\usepackage{amsmath}	% Advanced maths commands

\usepackage[dvipsnames]{xcolor} % remove later

\newcommand{\swift}{\textit{Swift}}

\title[Prompt LOFAR observations of GRB 210112A]{A LOFAR prompt search for radio emission accompanying X-ray flares in GRB 210112A}

\author[A. Hennessy et al.]{A. Hennessy$^{1}$,\thanks{E-mail: ah724@leicester.ac.uk}
R. L. C. Starling$^{1}$, 
A. Rowlinson$^{2,3}$, 
I. de Ruiter$^{2}$, 
A. Kumar$^{4}$, 
R. A. J. Eyles-Ferris$^{1}$, \newauthor 
A. K. Ror$^{5}$,
G. E. Anderson$^{6}$,
K. Gourdji$^{7,8}$,
A. J. van der Horst$^{9}$,
S. B. Pandey$^{5}$,
T. W. Shimwell$^{3,10}$,\newauthor
D. Steeghs$^{4}$,
N. Stylianou$^{1}$\thanks{Present address: Department of Physics, University of Oxford, Denys Wilkinson Building, Keble Road, Oxford OX1 3RH, UK},
S. ter Veen$^{3}$, 
K. Wiersema$^{11,12}$, 
R. A. M. J. Wijers$^{2}$\\
\\
$^{1}$School of Physics and Astronomy, University of Leicester, University Road, Leicester LE1 7RH, UK\\
$^{2}$Anton Pannekoek Institute for Astronomy, University of Amsterdam, Science Park 904, 1098 XH Amsterdam, The Netherlands\\
$^{3}$ASTRON, the Netherlands Institute for Radio Astronomy, Oude Hoogeveensedijk 4, 7991 PD Dwingeloo, The Netherlands\\
$^{4}$Department of Physics, University of Warwick, Coventry CV4 7AL, UK\\
$^{5}$Aryabhatta Research Institute of Observational Sciences (ARIES), Manora Peak, Nainital 263002, India\\
$^{6}$International Centre for Radio Astronomy Research, Curtin University, GPO Box U1987, Perth, WA 6845, Australia\\
$^{7}$Centre for Astrophysics and Supercomputing, Swinburne University of Technology, Hawthorn VIC 3122, Australia\\
$^{8}$OzGrav: ARC Centre of Excellence for Gravitational Wave Discovery, Hawthorn VIC 3122, Australia\\
$^{9}$Department of Physics, George Washington University, 725 21st St NW, Washington, DC 20052, USA\\
$^{10}$Leiden Observatory, Leiden University, PO Box 9513, NL-2300 RA Leiden, The Netherlands\\
$^{11}$Centre for Astrophysics Research, University of Hertfordshire, Hatfield, AL10 9AB, UK\\
$^{12}$Physics Department, Lancaster University, Lancaster, LA1 4YB, UK\\
}

\date{Accepted XXX. Received YYY; in original form ZZZ}

\pubyear{2023}

\begin{document}
\label{firstpage}
\pagerange{\pageref{firstpage}--\pageref{lastpage}}
\maketitle

% Abstract of the paper
\begin{abstract}
The composition of relativistic gamma-ray burst (GRB) jets and their emission mechanisms are still debated, and they could be matter or magnetically dominated. One way to distinguish these mechanisms arises because a Poynting flux dominated jet may produce low-frequency radio emission during the energetic prompt phase, through magnetic reconnection at the shock front. We present a search for radio emission coincident with three GRB X-ray flares with the LOw Frequency ARray (LOFAR), in a rapid response mode follow-up of long GRB 210112A (at $z\sim2$) with a 2 hour duration, where our observations began 511\,s after the initial \swift-BAT trigger. Using timesliced imaging at 120--168\,MHz, we obtain upper limits at $3\sigma$ confidence of 42\,mJy averaging over 320 second snapshot images, and 87\,mJy averaging over 60 second snapshot images. LOFAR's fast response time means that all three potential radio counterparts to X-ray flares are observable after accounting for dispersion at the estimated source redshift. Furthermore, the radio pulse in the magnetic wind model was expected to be detectable at our observing frequency and flux density limits which allows us to disfavour a region of parameter space for this GRB. However, we note that stricter constraints on redshift and the fraction of energy in the magnetic field are required to further test jet characteristics across the GRB population.
\end{abstract}


%%%%%%%%%%%%%%%%%%%%%%%%%%%%%%%%%%%%%%%%%%%%%%%%%%

%%%%%%%%%%%%%%%%% BODY OF PAPER %%%%%%%%%%%%%%%%%%

\section{Introduction}

    GRBs show a broad range of spectral and temporal behaviours, but are generally classified into two main categories. GRBs are defined by their spectral hardness and $T_{90}$ (the time for 5\% to 95\% of photons to be detected). Long gamma-ray bursts (LGRBs) are spectrally softer than their short gamma-ray burst (SGRB) counterparts, and have $T_{90} \gtrsim 2$\,s. LGRBs are the most common class \citep{Kouveliotou1993} and occur during the deaths of massive stars in core-collapse supernova, many have been associated with Type Ic supernova \citep{Galama1998,Hjorth2003}. SGRBs, on the other hand, are shorter in duration and are expected to occur during a binary neutron star (NS-NS) or neutron star-black hole (NS-BH) mergers \citep{Lattimer1976,Abbott2017}. This work presents observations of an LGRB.

    Relativistic jets are formed in the collapse \citep{Woosley1999}, and if pointed towards Earth, we observe an intense series of one or more $\gamma$-ray pulses lasting a few ms to a few hundred seconds. The pulses are attributed to internal shocks or magnetic reconnection events, close to the progenitor of the burst \citep{Rees1994}. The GRB jet cannot be directly resolved, and the composition and emission mechanisms still remain an open question in GRB physics.  There are models supporting both a matter-dominated \citep{Rees1994} or a magnetically dominated \citep{Drenkhahn2002} jet.
    
    As the jet propagates from the progenitor it will collide with intermediate matter and produce a unique afterglow. The properties of these are described by synchrotron, inverse Compton or synchrotron self-Compton emission, which results in broken power law behaviour both temporally and spectrally, these have been observed from X-ray to GHz radio \citep{Costa1997,vanParadijs1997,Frail1997}. GeV emission is also attributed to the afterglow \citep{Abdo2009}. The \textit{Neil Gehrels Swift Observatory} \citep[hereafter \swift,][]{Gehrels2004} detects X-ray afterglows in 95\% of GRBs it detects \citep{Evans2009}.

    Flares are often observed superimposed on the X-ray afterglow, usually in the first few hundred seconds. These flares show a variety of temporal features similar to the prompt emission \citep{Guidorzi2015}. X-ray flares and $\gamma$-ray pulses can be seen coincident in a number of bursts; and many studies favour that both have the same internal origins \citep[e.g.][]{Falcone2007,Chincarini2007,Hu2014}. There are a number of theories on the exact origins of these flares, including the late internal-shock model \citep{Fan2005}, late time energy injection \citep{Nousek2006} and misaligned jets \citep{Duque2022}. Pulses and flares have been shown to share the same power law distribution of peak flux, duration and waiting time as Solar flares, suggesting both may form through a similar magnetic process in very different environments.
    
    The magnetic reconnection process is capable of releasing quick, powerful bursts of energy, making it a suitable candidate to power GRB prompt emission. It has also been shown capable of producing the inherent variability in light curves \citep{Beniamini2016}. In this paper, we aim to test a magnetically dominated jet model. At least one magnetic model produces a testable prediction of short pulses of coherent radio emission corresponding to $\gamma$-ray pulses, but observed with a delay due to dispersion \citep{Taylor1993}. Electromagnetic radiation is spread and slowed as it propagated through the intergalactic medium (IGM); and interstellar medium in the host galaxy and the Milky Way. Longer wavelengths are affected more, thus we see radio signals at a delay compared to X-ray. Radio emission should occur at the shock front of a relativistic, magnetised wind as it interacts with ambient media, with a peak frequency at most a few MHz, well below the capabilities of current facilities. However, radio facilities such as the LOw Frequency ARray \citep[LOFAR,][]{vanHaarlem2013} can still detect the high frequency tail of this emission. LOFAR's rapid response is not fast enough to observe most prompt $\gamma$-ray flares, rather \citet{Starling2020} proposed a probe of the more accessible late-time X-ray flares. These occur on timescales typically reachable with the LOFAR rapid response mode capability of $\sim$5\,min. They also show that of a sample of 200 flares from 81 \swift\ GRBs, 44\% of flares would have been observable at 144\,MHz using LOFAR's rapid response mode ($\le$ 5 min) if generated in a magnetically dominated wind as described by \citet{Usov2000}. These flares occur in both long and short GRBs so we may expect to see these signals in all GRBs that have the correct flare characteristics. Demonstrating that radio signals produced analogous to prompt $\gamma$-ray pulses and X-ray flares, and hence directly linked to central engine activity, would provide great evidence for a Poynting flux dominated emission mechanism in GRBs.
    
    The radio signals in aforementioned model are powered by the same central engine activity that produces currently observable $\gamma$-ray pulses or X-ray flares. Demonstrating that radio signals produced analogous to these features, and hence directly linked to central engine activity, would provide great evidence for a Poynting flux dominated emission mechanism in GRBs.

    Coherent radio signals are also predicted as persistent emission and short pulses during the formation of a magnetar \citep{Zhang2001,Rowlinson2019a}. Some models predict SGRBs to form a magnetar in the burst event - emission is expected during spin-down and subsequent collapse into a black hole \citep{Zhang2014}. This may also apply for LGRBs, if a magnetar can be formed \citep{Bernardini2015}. The short pulses are likely to look like, and may contribute to, the population of Fast Radio Bursts (FRBs), another class of transient event characterised by short pulses of radio emission. These models postulate that at least some FRBs may occur from SGRBs \citep{Totani2013,Gourdji2020}. A possible ($2.8\sigma$) associated of an NS-NS merger and FRB was recently reported through a comparison between public gravitational wave and CHIME FRB data, where the FRB could be attributed to the magnetar collapse \citep{Moroianu2023}. An association would explain the origin of at least some FRBs, and would be a powerful probe of neutron star mergers.

    There have been a number of previous searches for prompt radio emission from GRBs, but often due to a lack of sensitivity or response time, it has proven difficult to obtain a detection \citep{Dessenne1996,Koranyi1995,Anderson2018}.
    
    Recent follow-up observations using LOFAR looked to test models of the formation of a magnetar. \citet{Rowlinson2019b} followed-up an LGRB beginning 4.5 minutes after the event, finding a $3\sigma$ upper limit of 1.7\,mJy\,beam$^{-1}$ for the full 2 hour 120--168\,MHz observation, and finding no detection of burst-like emission in timesliced data with cadence ranging from 10 minutes down to 30 seconds. This is the deepest low frequency limit to date, to our knowledge. A similar 120--168\,MHz follow-up of an SGRB finds a $3\sigma$ limit of 153\,mJy on emission during the plateau phase with 136 second integrated images \citep{Rowlinson2021}. This limit is two orders of magnitude greater than the predicted radio flux at the putative host redshift of 1.8, and they constrain the efficiency of rotational energy conversion into coherent radio emission to be $\le 6 \times 10^{-8}$.
    
    The Murchison Widefield Array \citep[MWA,][]{Tingay2013,Wayth2018}, a low-frequency radio facility in the southern hemisphere, whilst not able to reach the same flux limits as LOFAR, has the advantage of a very rapid response of $\sim$20\,s to a burst trigger, allowing it to probe the earliest times after a burst. A followup of an LGRB found no detection of pulse-like emission, acquiring fluence upper limits of 77--224\,Jy\,ms corresponding to 0.5--10\,ms pulse widths, some of the most stringent to date \citep{Tian2022b}. A search for emission analogous to X-ray flares was also conducted and the non-detection helped establish constrains on the fraction of magnetic energy. Investigations of short GRBs with MWA also have so-far found no detection of radio emission \citep{Kaplan2015,Anderson2021,Tian2022a}.
    
    Here we present a search for prompt coherent radio emission coincident with high energy flaring in long GRB\,210112A. This burst was observed as part of a LOFAR High Band Antenna (HBA) rapid-response campaign, and is a promising candidate for testing magnetic wind modes due to several prominent and simultaneous $\gamma$-ray pulses and X-ray flares, continuing to a time accessible by LOFAR (i.e. >5 minutes after burst). Section \ref{sec:observations} details the observations with \swift, LOFAR and ground-based optical telescopes used in this study. In Section \ref{sec:Xflares} we model the X-ray flares and place GRB\,210112A in the context of the \swift\ GRB afterglow sample. Section \ref{sec:radiopredictions} outlines the magnetic wind model which these observations can test, and we make predictions for the observable flux density at 144\,MHz with LOFAR. We present the results of our radio search in Section \ref{sec:radioresults}, and in Section \ref{sec:discussion} we examine the assumed redshift, discuss the implications of our findings on jet and progenitor models and look to future LOFAR upgrades, concluding in Section \ref{sec:concl}. Errors are stated as $1\sigma$ confidence unless otherwise stated.
    
\section{Observations and data reduction}
    \label{sec:observations}

    \begin{figure*}
        \includegraphics[width=\textwidth]{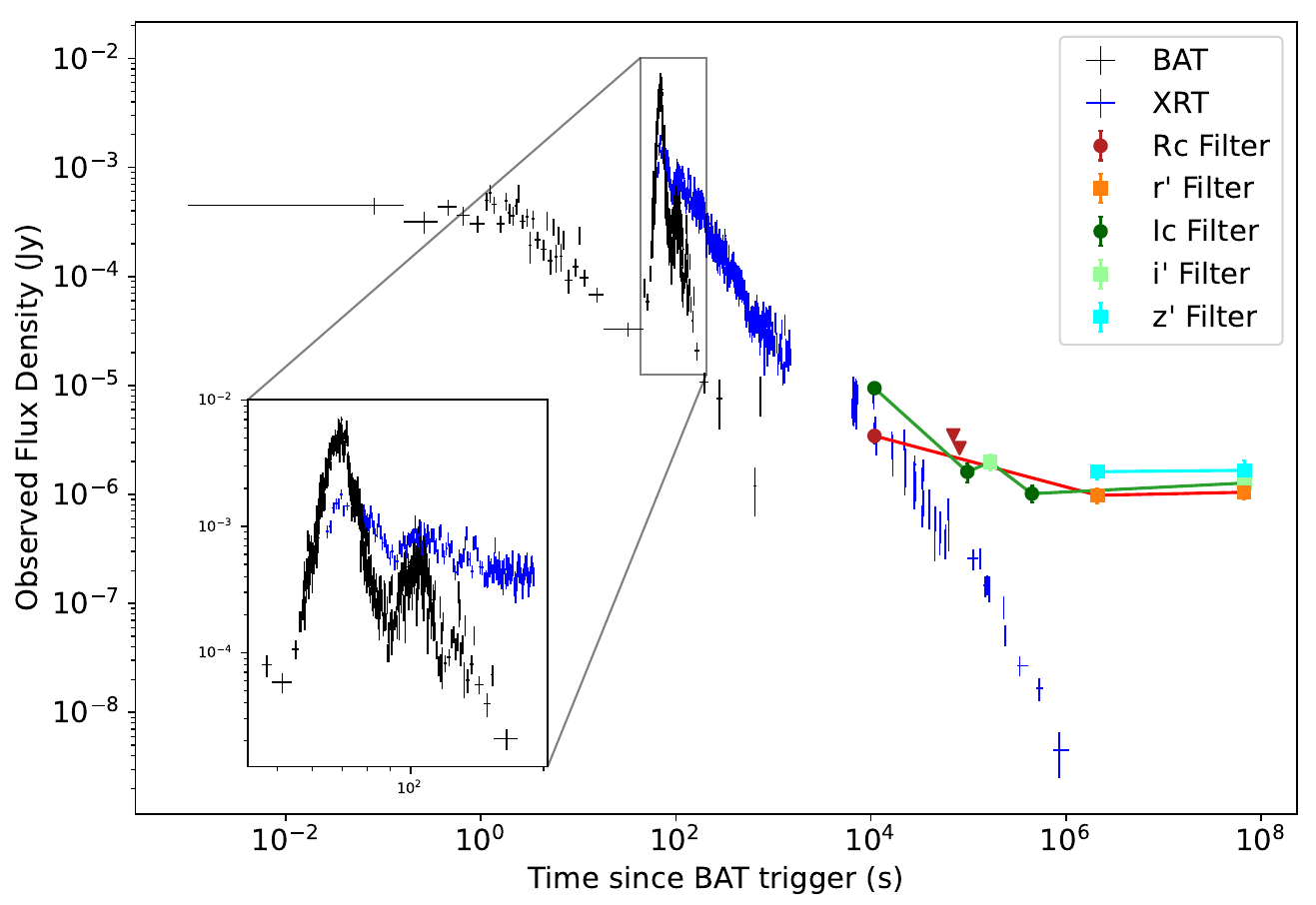}
        \caption{The observed flux density light curve of all \swift-BAT (at 50\,keV) and \swift-XRT (at 1\,keV) data, and ground-based optical data used in this paper, showing coincident $\gamma$-ray pulses and X-ray flares. The black datapoints are BAT, and blue represents the XRT data. Optical data (green, orange, purple) come from OSN, Mondy, DFOT, CAHA, DOT and LBT. See Section \ref{sec:observations} for references and \autoref{tab:opticaldata} for full optical data. Inset: a zoom on the main pulsing activity between T+40 and T+190\,s.}
        \label{fig:joint_lc}
    \end{figure*}

    GRB\,210112A occurred on the 12\textsuperscript{th} January 2021 at 01:51:00 GMT, triggering \swift\ \citep{gcnAmbrosi2021} and several other gamma-ray observatories (AGILE, \citealt{gcnUrsi2021}; Konus-Wind, \citealt{gcnSvinkin2021}; Insight-HXMT, \citealt{gcnZheng2021}). The burst fits into the LGRB category with a $T_{90}$ of $107.6 \pm 13.0$\,s in the 15--350\,keV energy range \citep{gcnStamatikos2021}. 

    Follow-up was conducted at X-ray \citep{gcnGoad2021,gcnEvans2021}, ultraviolet and optical wavelengths \citep{gcnSiegel2021} by \swift\ and a number of ground-based optical facilities: Kitab \citep{gcnNovichonok2021}, OSN \citep[Sierra Nevada Observatory,][]{gcnKann2021a, gcnKann2021b}, DFOT \citep[Devasthal Fast Optical Telescope,][]{gcnGupta2021}, Mondy \citep{gcnPankov2021}, CAHA \citep[Calar Alto Observatory,][]{gcnKann2021c}, DOT \citep[Devasthal Optical Telescope,][]{gcnDimple2021, gcnMisra2021}, LBT \citep[Large Binocular Telescope,][]{gcnRossi2021}. Our rapid-response LOFAR observations were automatically triggered, and LOFAR HBA began observing the target shortly after the X-ray position was reported, the GRB position being immediately available for LOFAR, 511 seconds after the burst.
    
    Subsequent sections detail the observations and methods of analysing the data. The $\gamma$-ray, X-ray and optical data used in this work are shown together in \autoref{fig:joint_lc}. BAT and XRT are plotted at 50\,keV and 1\,keV, respectively \citep{Evans2010}. Optical data are plotted at the central frequency of each filter, accounting for a Galactic extinction according to \citep{Schlafly2011}.
    
    \subsection{\swift}

    The Burst Alert Telescope \citep[BAT,][]{Barthelmy2005} on board \swift\ was triggered at 01:37:03 UT, 12\textsuperscript{th} January 2021. On detection, the observatory slewed and the X-ray Telescope \citep[XRT,][]{Burrows2005} began observing 74.2\,s after the initial BAT trigger, finding an uncatalogued X-ray source at RA, DEC (J2000) = 14h 36m 1.06s, +33\degr 03\arcmin 23.8\arcsec\, (90\% error radius $1.4\arcsec$).
    83\,s after the initial trigger, the UV-Optical Telescope \citep[UVOT,][]{Roming2005} began observing and detected a source in the White filter with a 147\,s exposure. The detection was accumulated in an image spanning more than the duration of the prompt emission, so is not utilised further. The {\it V, B, U}, and {\it UVW1} filters were also used at later times, but no source was detected in these images and only upper limits were derived \citep{gcnSiegel2021}.

    We obtained the data from the online \swift\ archive\footnote{\url{https://www.swift.ac.uk/swift_portal/}} at the UK Swift Science Data Centre \citep[UKSSDC, see][]{Evans2009}. Data reduction and spectral analysis were carried out in {\sc Heasoft}\footnote{\url{http://heasarc.gsfc.nasa.gov/ftools}} \citep[version 6.31.1,][]{Heasoft2014} and {\sc Xspec} \citep[version 12.13,][]{Arnaud1996}.
    
        \subsubsection{\swift-BAT}

        The light curve presented in \autoref{fig:joint_lc} demonstrates several $\gamma$-ray pulses over the observation. The spectral data were extracted with the {\sc batbinevt} pipeline. The data were timesliced according to visually identified pulses and availability of overlapping XRT data, described in \autoref{tab:spectralfits}. The BAT spectral data were modelled with a cutoff power law, where the photon index, exponential cutoff and normalisation were free parameters.

        \subsubsection{\swift-XRT}
        \label{sec:xrt_observations}
        
        The light curve uses data from all three modes of \swift\ operation: slewing, windowed timing and photon counting.

        The spectral data in windowed timing and photon counting modes were timesliced as defined by \autoref{tab:spectralfits}, beginning at the decay of the first flare. Each timeslice was fitted with one of several models. 

        For periods where there is no overlap between BAT and XRT data, a power law with two components of absorption was fitted \citep{Wilms2000}. The first component models the line-of-sight absorption through our own Galaxy with the column density calculated from \citet{Willingale2013} based on the location of the GRB, and this remains fixed across the full time interval of the GRB. For the position of GRB\,210112A, this gives a value of $N_{\rm H,Gal} = 9.55\times10^{19}\,\textrm{cm}^{-2}$. The second absorption parameter is an intrinsic component due to absorption in the line-of-sight IGM and host galaxy of the GRB source. This should also remain constant through the observation, as the parameter is not directly linked to the burst itself. Thus, we first create a late time spectrum ($T$+8913 to $T$+$1.7\times10^{5}$) with this parameter allowed to vary. We select late time data as the burst is less energetic at this point - the timescale and magnitude of variability is smaller in the afterglow. We find a value of $N_{\rm H,host} = 2.98\,(\pm\,0.33)\times10^{22}\,\textrm{cm}^{-2}$ (90\% confidence interval) and freeze this for all other spectral fits. A tentative redshift value of $z\sim2$ is reported in \citet{gcnKann2021a} which we include as a frozen value, though we acknowledge it is an estimate. We revisit this value in Section \ref{sec:discussion}. For periods of overlapping BAT and XRT data, we fit the absorbed broken power law model, with absorption described as above, and a spectral energy break allowed to vary, which in all three joint fits remained at $\sim13.5$\,keV.

\begin{figure}
        	\includegraphics[width=\columnwidth]{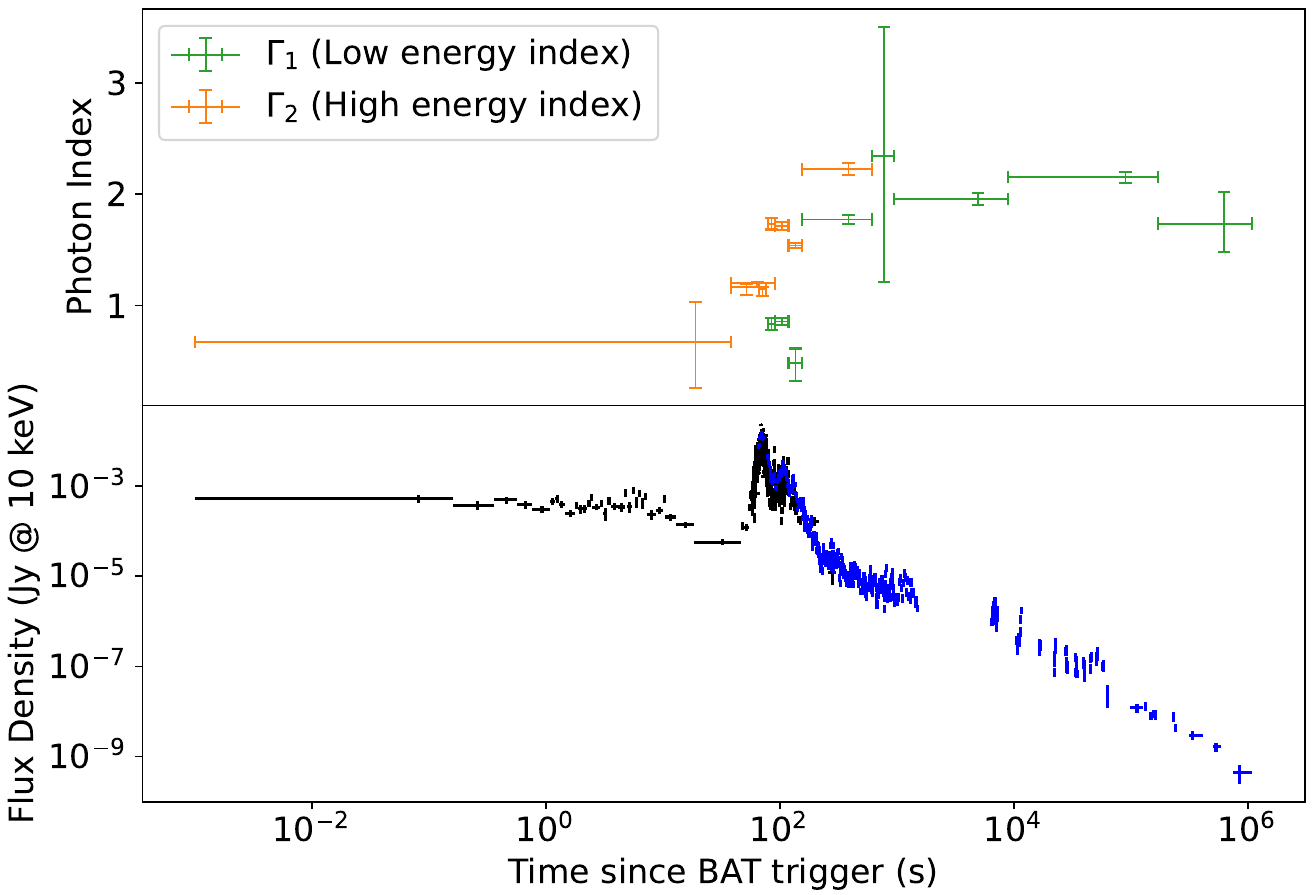}
            \caption{The spectral evolution of the power law index parameters, for each timeslice of data. Where there is combined data, $\Gamma_{1}$ is the low energy photon index and $\Gamma_{2}$ the high energy counterpart. Otherwise, they are the BAT fitted power law and XRT fitted power law, respectively. The full set of timeslices and fit parameters are tabulated in \autoref{tab:spectralfits}. Error in photon index is the fitted parameter 90\% confidence interval. Bottom: combined BAT and XRT flux density (extrapolated to 10 keV) light curve, for reference.}
            \label{fig:spectral_pars}
        \end{figure}

        The spectral evolution of power law indices is shown in \autoref{fig:spectral_pars}. The low energy and the high energy indices represent the spectral indices before and after the energy break, respectively, and where there is not a joint model, simply represent the spectral index of the fitted XRT or BAT data. Both spectral indices evolve as expected in a GRB - the spectrum is initially hard during the energetic initial stages of the burst, but rises to $\Gamma \sim 2$ during the afterglow. 

    \subsection{LOFAR}

        At 01:45:35 UTC, LOFAR triggered on GRB\,210112A in Rapid Response Mode obtaining 7199\,s of observation on target, starting 511\,s after the initial \swift-BAT trigger (project code LC15\_013). Observations were made with LOFAR HBA that operate in the 120--168\,MHz frequency range. The data were averaged with a 1 second integration time and 244 sub-bands with a bandwidth of 195.3\,kHz (16 channels per sub-band). 23 core stations and 11 remote stations were used (the Dutch array). After the target observations were completed, a 10 minute calibrator observation was taken using 3C147 with the same time and frequency averaging.
        
        We calibrated the data using the {\sc LINC}\footnote{\url{https://git.astron.nl/RD/LINC}} pipeline (version 4.0), a pipeline to correct for instrumental effects in LOFAR observations, with 8 second time averaging. Standard LOFAR software and methods are used as described in \citet{deGasperin2019,vanWeeren2016} and \citet{Williams2016}. The pipeline uses the calibrator data to derive direction independent gain solutions. These solutions are applied to the target data and an initial step of phase-only calibration is applied using a sky model from TGSS ADR \citep{Intema2017}. Part of the pipeline includes {\sc AOFlagger} \citep[version 3.2.1,][]{Offringa2010, Offringa2012}, enabling statistical flagging and removal of data affected by radio interference. Further direction dependent calibration was not required in our case, as the target GRB position lies at the centre of the observation field.

        The calibrated data were imaged using the {\sc WSClean} \citep[version 3.1.1,][]{Offringa2014}. A large-scale, deep image was created using the full time, full frequency range observation. A 5\,arcsec pixel scale was used, and standard imaging parameters included a Briggs weighting -0.5 and auto-thresholding up to 100,000 iterations. Timesliced images were created for 60 second and 320 second length time bins, with the same imaging parameters as before, using the full frequency range and a 1 arcsecond pixel scale. Subtraction imaging was attempted by subtracting the calculated model visibilities from the target data, but we saw negligible improvement on the rms noise of the images.

    \subsection{Optical Observations}
    \label{sec:optical_observations}

    An optical counterpart was detected in several ground-based facilities - see \autoref{tab:opticaldata}. Using optical follow-up data collected 3 hours after the GRB with the Sierra Nevada Observatory (OSN), \citet{gcnKann2021a} fit an SED consistent with a large amount of Milky Way like dust at $z\sim2$.

    A flattening in the optical light curves was noted in GCN Circulars and attributed to the host galaxy \citep{gcnKann2021c,gcnRossi2021}. To confirm the host galaxy magnitudes, we obtained further photometry of the GRB\,210112A field. The observations were carried out using the ARIES-Devasthal Faint Object Spectrograph and Camera \citep[ADFOSC,][]{Omar2019} mounted at the 3.6m Devasthal Optical Telescope \citep{Kumar2018}. 
    
    Imaging observations in the SDSS {\it r'} and {\it i'} bands ($300{\rm s}\times10$ frames and $300{\rm s}\times10$ frames, respectively) were taken on 2023-02-21, and observations in the {\it z'}-band ($300{\rm s}\times15$ frames) were taken on 2023-03-27; more than two years since the burst. 
    
    Pre-processing of the raw data (bias, flat, and cosmic-ray correction) was carried out according to standard procedure. The multiple frames observed on the same night were aligned and stacked to achieve a good signal-to-noise ratio. The host galaxy of GRB\,210112A is detected clearly within the 3-$\sigma$ limit in the stacked frames of {\it r', i'}, and {\it z'} bands.

    The photometric analysis is conducted using Source Extraction and Photometry \citep[SEP,][]{Barbary2018}, a python Package built on the Source-Extractor \citep[SExtractor,][]{Bertin1996} software suitable for performing faint-galaxy photometry, and SDSS standard stars were used to perform the calibration. The estimated calibrated {\it r'}, {\it i'}, and {\it z'} band magnitudes of the host galaxy of the GRB\,210112A are $23.89 \pm 0.12$, $23.67 \pm 0.24$ and $23.35 \pm 0.25$ AB mags, respectively, before accounting for the expected Galactic extinction of $E(B-V) = \sim0.01$ \citep{Schlafly2011}.

\section{X-ray flare analysis}
\label{sec:Xflares}

    In order to generate predictions for prompt radio emission, discussed later in Section \ref{sec:radiopredictions}, we need to calculate the fluences of the X-ray flares. We have developed a code that can model GRB temporal behaviour as a power law afterglow plus flares. An outline of this process is described in the subsequent section, but will be detailed fully in a future paper.

    \subsection{Light curve fitting procedure}
        \label{sec:laff}

        The program uses light curve data made available on the Swift Online Archive. The initial step is to identify potential flares, before refining these into a set of start, stop and peak times. Possible flares are identified by looking for trends where the flux increases in at least 2 out of 3 consecutive data points from point $n$ - additionally each increase must be by a factor of greater than twice the point $n$'s 90\% confidence interval. This criterion is somewhat arbitrary, but shows consistent success upon visual inspection. For each potential flare, we find the start and peak by finding the local minima and maxima, respectively. Identifying the end of the flare is a more involved process as it fades into the afterglow. We based our method on \citet{Evans2009}. For each point we calculate two decay indices, $\alpha_\textrm{peak}$, the gradient from the flare peak to point $n$; and $\alpha_\textrm{next}$, the gradient from point $n$ to $n+1$. For the decay to end, we require two conditions to be met at least 6 times over any 10 consecutive points. Condition 1 requires that $\alpha_\textrm{peak}>\alpha_\textrm{next}$, the gradient from $n$ to $n+1$ is shallower than gradient from peak to $n$. Condition 2 requires that $\alpha_\textrm{peak}$ and $\alpha_\textrm{next}$ are greater in bin $n$ compared to bin $n-1$, both indices are shallowing over consecutive points.
        
        With a set of flares identified, we exclude those data such that we are left with only the afterglow continuum. A series of broken power laws are fitted, each with a different number of breaks up to 5, to cover a range of common afterglow light curve shapes seen in GRBs \citep{Nousek2006,Zhang2006,O'Brien2006}. A least squares method is used to fit the data. For each model, we introduce the Akaike Information Criterion \citep[AIC,][]{Akaike1974}, an estimator of relative model quality. This allows us to use a penalty function to avoid over fitting the continuum. For each fit, we calculate:

        \begin{equation}
            \label{eqn:AIC}
            \Delta\textrm{AIC} = 2k + N\,\textrm{ln}(\textrm{RSS})\,,
        \end{equation}

        where $k$ is the number of parameters modelled, $N$ the number of data points, RSS is the residual sum of squares, and $\Delta\textrm{AIC}$ is the quantity we look to minimise.
        
        We re-add the flare data and subtract the fitted continuum - leaving only residuals and clear flare peaks. For each flare, we fit a fast-rise, exponential-decay (FRED) curve, commonly found to be a good representation of GRB pulses \citep{Norris1996,Peng2010} From this we can find the temporal parameters for each flare, and more usefully, we can integrate across the flare model to derive the X-ray fluence for each flare.
    
    \subsection{Flares in GRB\,210112A}
     \label{sec:laff_results}

\begin{figure}
            \includegraphics[width=\columnwidth]{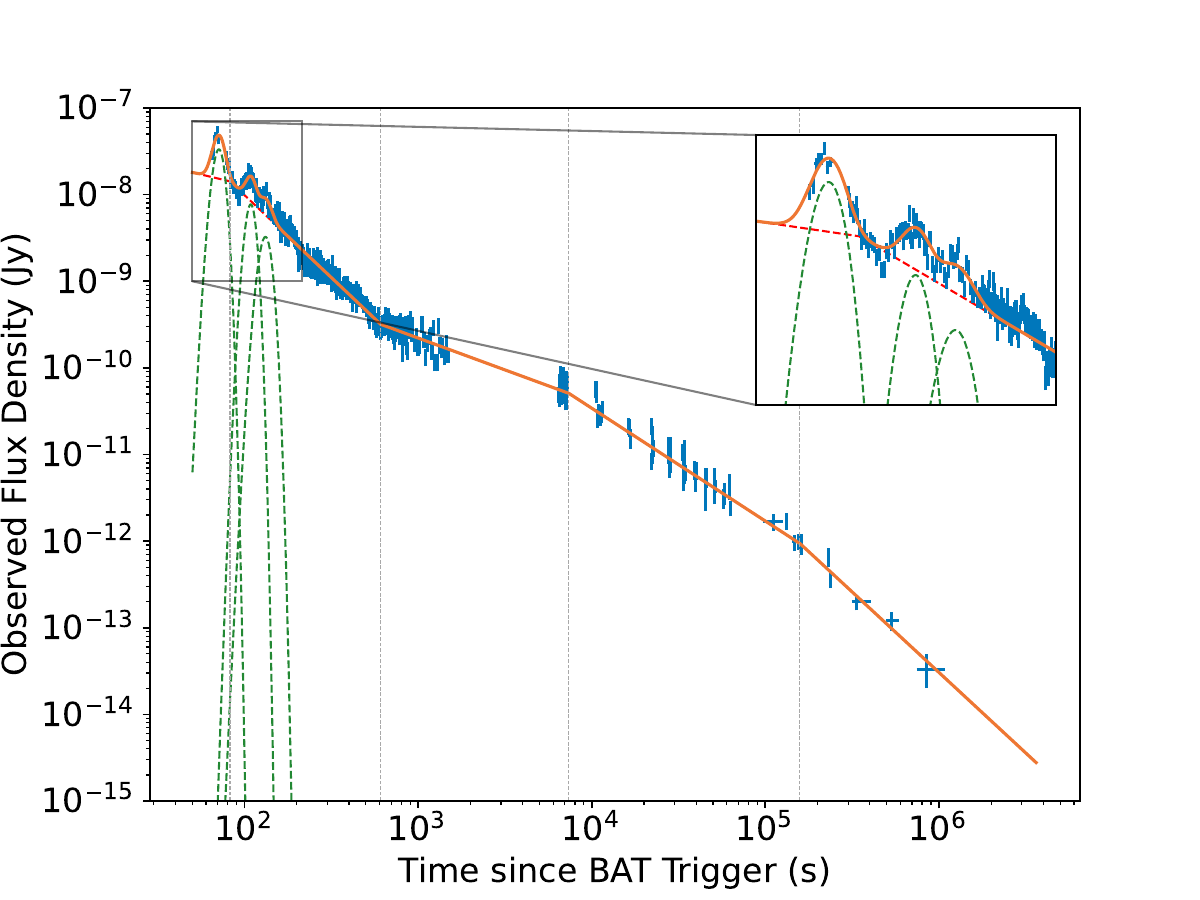}
            \caption{Observed XRT flux density light curve of GRB\,210112A are shown in blue, fitted using our code described in Section \ref{sec:laff}. Orange line represents the fitted model; red and green dashed lines are the continuum and flare components, respectively. Grey vertical lines show the power law breaks. Inset: zoom at 40--320 seconds showing the early X-ray flaring.}
            \label{fig:laff_fit}
        \end{figure}

        The light curve fitted with our code is shown in \autoref{fig:laff_fit} and the full set of fit parameters are shown \autoref{tab:temporalfits}. The burst appears to look like a standard "canonical" light curve, showing all phases described in \citet{Zhang2006,Nousek2006}. We find three flares, all occurring during the ongoing prompt emission phase, at times 64.9--89.9\,s, 92.1--125.1\,s and 126.3--152.8\,s (relative to BAT trigger time). All three flares are observed to be coincident in X-ray and $\gamma$-ray, as shown in the joint light curve in \autoref{fig:joint_lc}, suggesting they are both formed by the same internal process - therefore we attribute the X-ray flares to the prompt emission phase. The X-ray fluence outputs for each flare are shown in \autoref{tab:modelpredictions}. The code also outputs a fluence error of $3.3\times10^{-9}$, $6.6\times10^{-9}$ and $6.8\times10^{-9}$ erg\,cm$^{-2}$ (0.3--10\,keV) for each respective flare, but these are small relative to the later discussed model uncertainties and do not translate to predicted radio flux errors.
        
        There is a possibility of a late time jet break at $t \sim 10^{5}$\,s, our method of an AIC penalty function prefers a model fit with a late-time jet break, and there is a hardening of the afterglow spectrum at this time, as presented in \autoref{fig:spectral_pars}. However, a F-test significance test shows this temporal jet break fit is only significant to a $\sim 1.5\sigma$ significance level, over a broken power law fit without this break. The spectral hardening at this time is also significant only to $\sim 1.65\sigma$.
    
        GRB\,210112A is consistent with the population of LGRBs when plotted on the hardness-duration distribution \citep{Kouveliotou1993} and the Amati relation \citep{Amati2006}. Taking values from the set of spectral and temporal fits across the evolution, they are consistent with the distribution of parameters of \swift-XRT detected GRBs in \citet{Evans2009}, and this burst is typical in terms of $N_{\textrm{H}}$, break times, spectral and temporal indices. Additionally, the properties of each flare are consistent with the population of \swift\ X-ray flares in \citet{Yi2016}.
    
        All decay indices and spectral indices are consistent with pre-jet break expectations, taking the end of the XRT data as the earliest possible time for a jet break, we can derive constraints, for $z=2$, on the jet opening angle $\theta_j$ and total energy budget $E_\gamma$ of this burst. Following the methods of \citet[and references therein]{Starling2009}, we derive $E_{\textrm{iso}} = 3.62 \times 10^{52}$\,erg, and lower limits of $\theta_{j} \ge 0.143$\,rad and $E_{\gamma} \ge 3.71 \times 10^{50}$\,erg; for a flat universe with $H_{\textrm{0}} = 67.4$ km s$^{-1}$ Mpc$^{-1}$, $\Omega_{\textrm{M}} = 0.315$ and $\Omega_{\textrm{vac}} = 0.811$ \citep{PlanckCollaboration2020}. These parameter limits are consistent with the population of jet break parameters shown in \citet{Zhao2020}.

\section{Search for coherent low-frequency radio emission}
\label{sec:radiopredictions}

    It is possible to make predictions for the expected radio flux densities if the emission originates in a magnetic wind model, as outlined in \citet{Usov2000}. The relativistic, strongly magnetised winds interact with the ambient medium to produce the observed synchrotron spectrum, as well as low-frequency electromagnetic waves.

    \subsection{Radio predictions}

    We test these predictions using LOFAR HBA data, with a central observing frequency of 144\,MHz and a bandwidth of 48\,MHz. The peak frequency of radio emission is
    \begin{equation}
        \label{eqn:vmax}
        v_{\textrm{max}} = \frac{1}{1+z}\epsilon_{B}^{1/2}\,\textrm{MHz}\,,
    \end{equation}
    where $z$ is the GRB redshift, and $\epsilon_{B}$ is the fraction of total energy in the magnetic field. This parameter is not well defined, and this is discussed later in \autoref{sec:discussion}, but we take $\epsilon_{B} = 10^{-3}$, following \citet{Katz1997}. For GRB 210112A, with our $z=2$ assumption, we find a peak frequency of 0.011\,MHz. This is well below the capability of current radio facilities, however, LOFAR is sufficient to probe the high energy tail of this emission.

\begin{figure}
        \includegraphics[width=\linewidth]{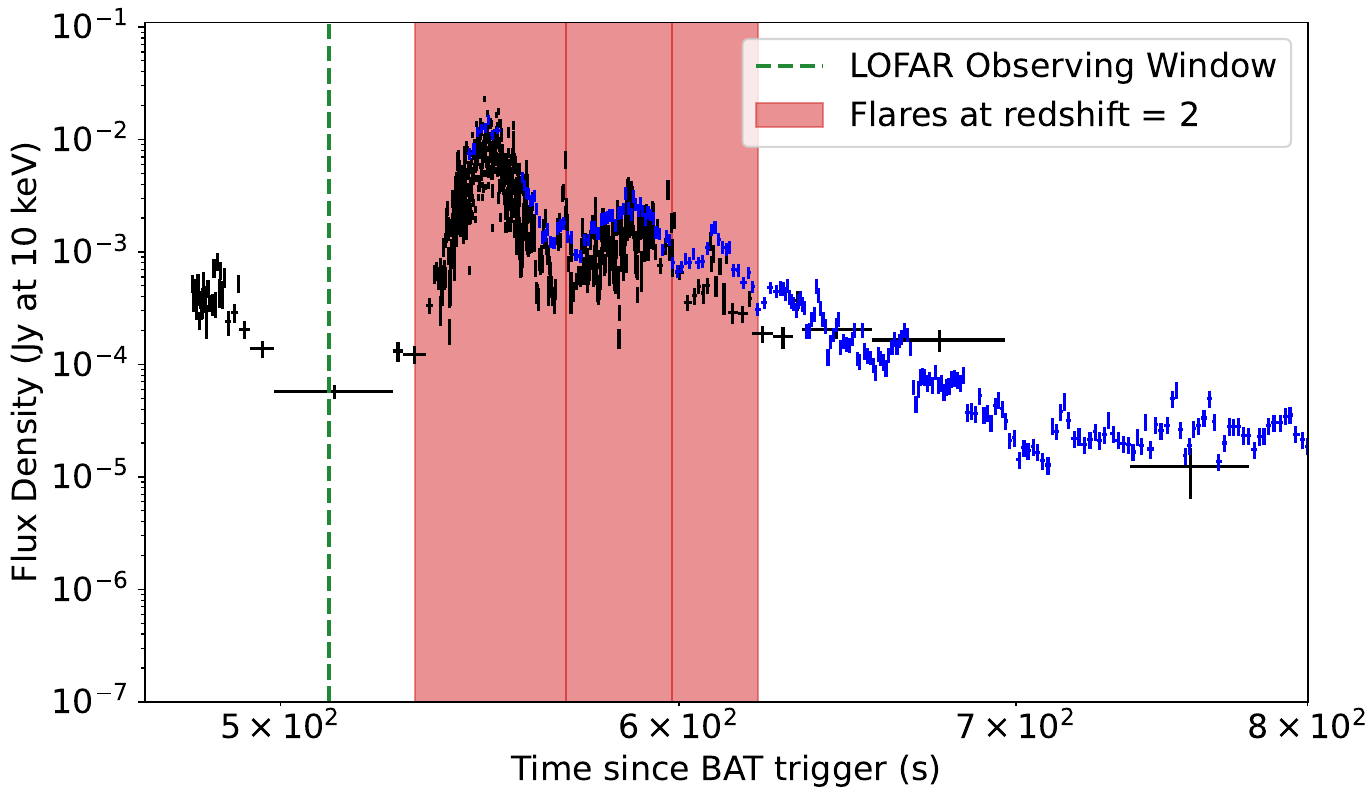}
        \caption{The BAT (black) and XRT (blue) light curve with a dispersion delay applied - for LOFAR 144\,MHz HBA observations at $z=2$, equivalent to a 480\,s shift. The green vertical line marks the start of the LOFAR observing window (the end is not displayed on this figure), and red regions mark the flares. At $z=2$, the dispersion delay is such that all three flares and the plateau lie within the LOFAR observing window.}
        \label{fig:radioflares}
    \end{figure}

    The radio emission is delayed compared to the analogous prompt pulses, as a result of dispersion through the line of sight. From \citealp{Taylor1993} (see also \citealp{Cordes2003}), the dispersion delay is given by
    \begin{equation}
        \label{eqn:dispersion}
        \tau(\nu) \sim \frac{\textrm{DM}}{241\nu^2}\,\textrm{s}\,,
    \end{equation}
    where DM is the dispersion measure and $v$ the observing frequency in GHz. We expect DM to scale with redshift, but this relation is not well defined. In particular, the contribution of the environment near to the burst and host galaxy is unknown. For a more in-depth discussion of dispersion measure, see \citet{Macquart2020} and \citet{James2022}. In this case, we follow \citet{Lorimer2007} and estimate DM as $\sim1200z$\,pc\,cm$^{-3}$. Given this, we expect a delay $\tau(\nu) = 480$ seconds, placing all three flares in our LOFAR observation window. \autoref{fig:radioflares} shows the the $\gamma$-ray and X-ray data with the corresponding dispersion delay applied.

    The observed pulse width will be longer than the intrinsic pulse width after being spread through dispersion. As the pulse occupies a range of frequency, each part will be delayed by a differing amount, spreading the signal. If we model the temporal shape of the intrinsic pulse as a delta function, the observed duration $\tau_{r}$ can be estimated, as given in \citet{Usov2000}:
    \begin{equation}
        \label{eqn:duration}
        \tau_{r} \sim 2\frac{\Delta\nu}{\nu}\tau(\nu)\,\textrm{s}\,,
    \end{equation}
    where $\Delta\nu$ is the observing bandwidth in GHz and $\tau(\nu)$ the dispersion delay given in \autoref{eqn:dispersion}. The real intrinsic pulse will have some finite length so the observed radio pulse should actually be longer, the peak radio emission should occur in the time period $\tau_{r}$. 

    Finally, we can calculate the radio spectral flux density for a radio pulse, based on the X-ray fluence of the XRT flares. Given by \citet{Usov2000}, the flux in the dispersion limited and non-dispersion limited cases for each flare are given as:
    \begin{equation}
        F_{v} =\begin{cases}
        \frac{\delta(\beta-1)}{\tau_{r} v_{\textrm{max}}}\left(\frac{v}{v_{\textrm{max}}}\right)^{-\beta} \frac{\Phi_{\gamma}}{10^{-23}}\,\textrm{Jy}, & \tau_{r} \leq \frac{2\Delta v}{v}\tau(v)\\
        
        \frac{\delta(\beta-1)}{2\Delta v\tau}\left(\frac{v}{v_{\textrm{max}}}\right)^{1-\beta} \frac{\Phi_{\gamma}}{10^{-23}}\,\textrm{Jy}, & \tau_{r} > \frac{2\Delta v}{v}\tau(v)
        \end{cases},
    \end{equation}
    where $\delta$ is the ratio of bolometric radio fluence to bolometric $\gamma$-ray fluence; $\beta$ is the power law index in the high frequency radio tail; $\tau_{r}$ is the intrinsic flare length and $\Phi_{\gamma}$ is the bolometric $\gamma$-ray fluence. Following \citet{Starling2020} and references therein, we take $\delta \sim 0.1 \epsilon_{B}$ and $\beta \sim 1.6$. For the three X-ray flares identified in Section \ref{sec:laff_results}, we obtain the radio fluence predictions in \autoref{tab:modelpredictions}.

    Dense regions around the progenitor and in the line-of-sight may block emission where the plasma is opaque to the low frequency radio emission. This frequency is dependent upon the number density of electrons in the plasma. \citet{Zhang2014} estimate the co-moving plasma frequency in the shocked ejecta region of the blast wave as $\sim 4$\,MHz, which is comfortably below our observation band, so we expect the emission to escape along the jet axis. Another effect that can contribute to low frequency propagation is the interaction of low frequency radiation and electrons, leading to free-free absorption. Adopting a free-free turnover of 300\,MHz \citep{Lyutikov2016,Piro2016}, this would lie at 100\,MHz in our observed band for $z=2$, below the minimum observation frequency, and hence free-free absorption is not expected to heavily impact emission at 144\,MHz.

    \begin{table}
        \caption{The flux density of the predicted radio flares, for the three flares of GRB 210112A.}
        \label{tab:modelpredictions}
        \begin{tabular}{crr}
            \hline
            Flare No. & Observed Fluence & Predicted Flux Density\\ 
             & (0.3--10\,keV) (erg\,cm$^{-2}$)& at 144 MHz (mJy)\\
            \hline
            1 & $4.45\times 10^{-7}$ & 191\\
            2 & $1.29\times 10^{-7}$ & 56\\
            3 & $5.44\times 10^{-8}$ & 23\\
            \hline
        \end{tabular}
    \end{table}
    
    \subsection{Radio results}
    \label{sec:radioresults}

\begin{figure}
        \centering
        \includegraphics[width=\linewidth]{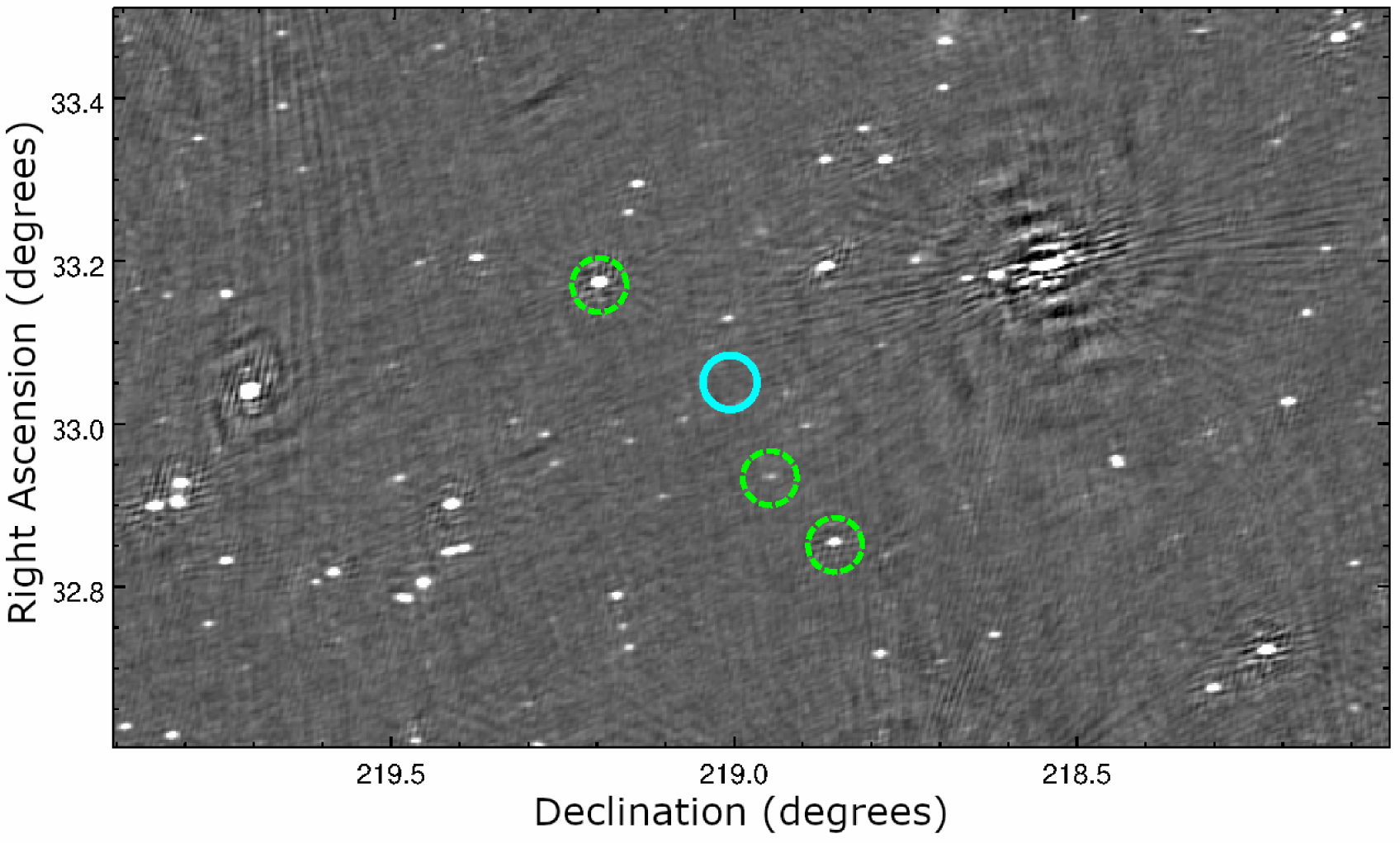}
        \caption{The full-time LOFAR image spanning the full 7199\,s of observation, showing GRB\,210112A and the surrounding region. The blue circle shows the position of the GRB and dashed green circles show a select few comparison sources. The size of the circles are 2 arcminutes and are meant only to guide the eye.}
        \label{fig:WideImage}
    \end{figure}

\begin{figure*}
        \centering
        \includegraphics[width=\textwidth]{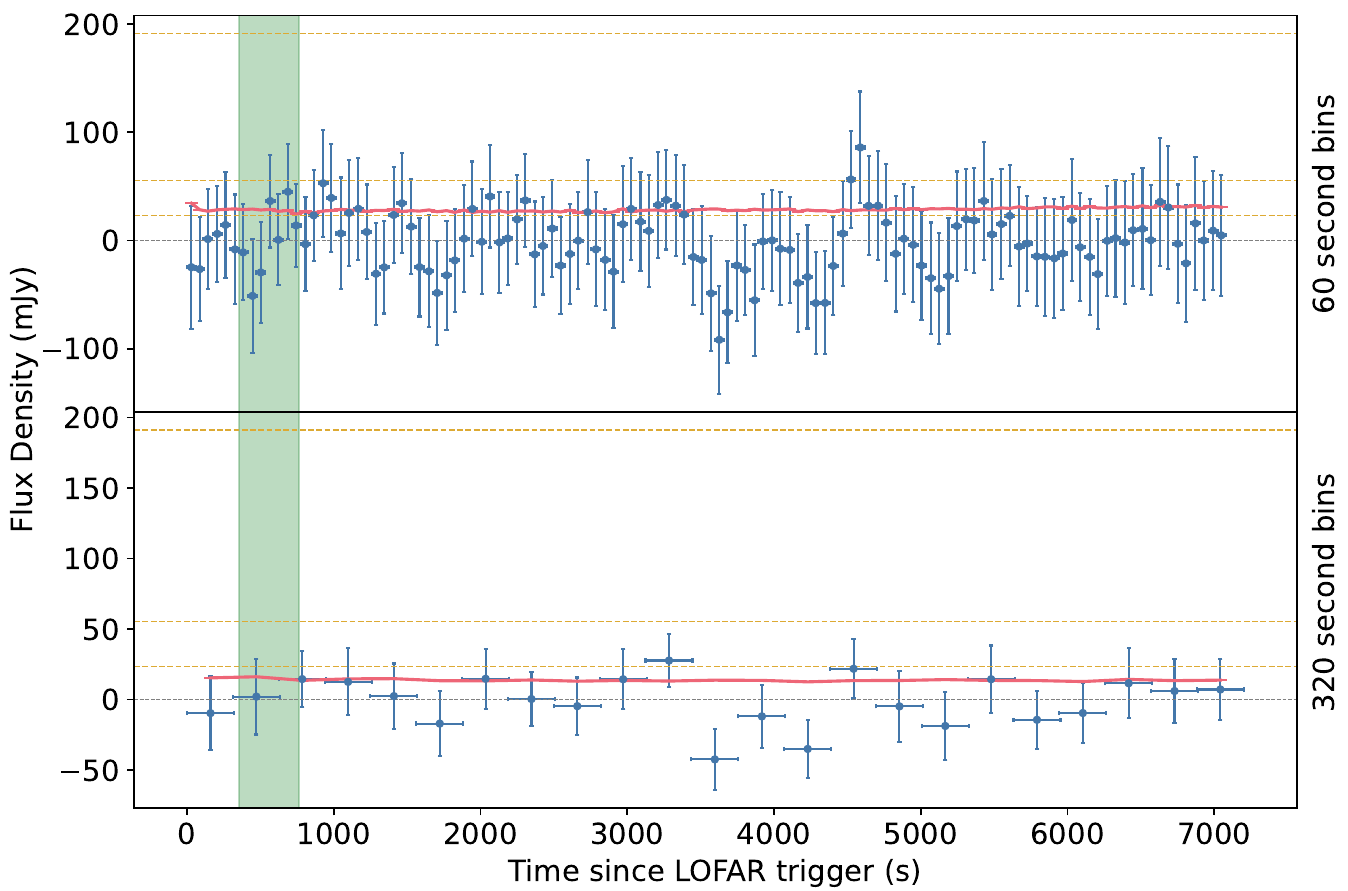}
        \caption{The observed 144\,MHz radio flux at the position of GRB\,210112A as a function of time. We show data from images created with 60 second bins and 320 second bins. The red line shows the $1\sigma$ rms noise measured in each image, calculated from the inner 1/8\textsuperscript{th} of the image. The yellow dotted lines shows the expected flux density values for the three X-ray flares predicted in Section \ref{sec:radiopredictions}. The green shaded region shows the expected time of arrival of all three radio pulses for $z=2$, the dispersed signals causes all three to overlap.}
        \label{fig:RadioLightcurve}
    \end{figure*}

    The large scale radio image generated from the full time 2 hour observation is shown in \autoref{fig:WideImage}, showing the surrounding region, position of our GRB and position of several bright comparison sources. Source extraction was performed using {\sc TraP} \citep[version 5.0,][]{Swinbank2015}, using a force fit extraction monitor at the position of our GRB and comparison sources, using default settings and restoring beam shape. These positions are monitored at each timeslice in our 60 second and 320 second snapshot images. \autoref{fig:RadioLightcurve} shows the resultant light curve of these snapshot images, marked with the expected arrival time of the radio flares. At the position of GRB 210112A, however, we do not detect significant radio emission.

    {\sc TraP} outputs an rms value, calculated with the inner default 1/8\textsuperscript{th} of the image, centered on the GRB. Given the non-detection of radio emission at our GRB position, we use this output rms value of 14\,mJy for our 320 second snapshot image. We therefore calculate the $3\sigma$ (3$\times$ rms) upper limit on any flaring activity at this integration length to be 42\,mJy. The flare may be split across multiple time bins, if the bin does not line up exactly with when the flare starts and ends, meaning the flare could be smeared out - at worst it is split evenly across two bins. \citet{Carbone2017} and \citet{Chastain2022} perform simulations and discuss how this could affect transient searches. However, in the case of the first flare at least, the radio flux should be bright enough to always be visible in each bin.

    Additionally, we image with 60 second snapshots which would catch a flare split across multiple bins, and to look for any shorter timescale variations in the data. In these images, we also detect no source at the expected position. We obtain an rms noise of 29\,mJy and consequently a $3\sigma$ upper limit of 87\,mJy.

    The predicted flux of the brightest radio flare lie well above the rms noise in the data for both time intervals, thus we can confidently say the flare predicted by the model would be seen provided the input assumptions are correct.

    We use the full 2 hour duration image to calculate a similar $3\sigma$ flux density limit. Any predicted flares are covered by integrating noise in these images, but this deep image allows us to put a limit on persistent emission from the region - for example from star formation within the host galaxy. For this, we yield an upper limit of 3\,mJy.
 
\begin{figure}
        \centering
        \includegraphics[width=\linewidth]{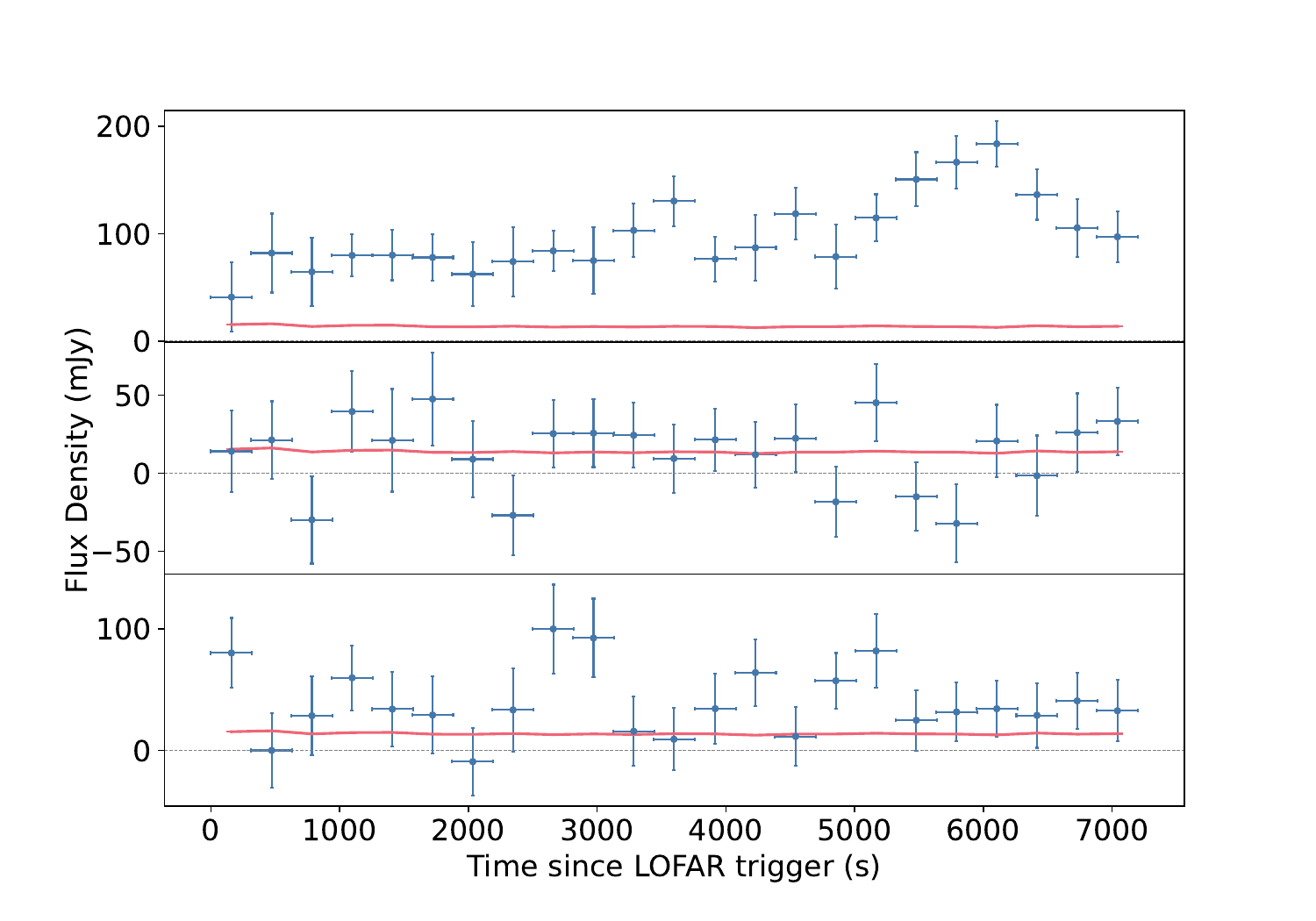}
        \caption{The observed 144 MHz radio flux at the position of several comparison sources in the same observation field as GRB\,210112A, as shown in \autoref{fig:WideImage}, as a function of time. Images are created with 320 second bins. Shown on the bottom two sub-figures are the rms noise in the images, in red. The top figure shows a source significantly brighter than the noise even in these snapshot images.}
        \label{fig:comparisonsources}
    \end{figure}

    We also measure the flux density for three comparison sources in the field indicated in the full images, using the same 320 second snapshot intervals. \autoref{fig:comparisonsources} shows we recover the brightest of three sources in all images. The two fainter comparison sources are not recoverable in time-sliced data, and this is consistent with the measured rms for this choice of binning.

\section{Discussion}
\label{sec:discussion}

    Our search for coherent radio emission has therefore result in no detections for any flares. The 3\,mJy limit on the GRB position obtained from the full two hour image may be reasonably compared to those for long GRB 180706A. Our 2 hour observation $3\sigma$ upper limit is 1.76 times higher than that found for GRB 180706A using the same instrument and approach \citep{Rowlinson2019b}. A forced fit at the position of GRB 180706A found a 1$\sigma$ flux density excess of $1.1\pm0.9$\, mJy, which would not be recoverable in the dataset of GRB 210112A, while it still represents one of the best constraints available at present.
    
    The sensitivity of our observations was sufficient to have seen the first predicted radio flare described by the \citet{Usov2000} magnetic wind model, analogous to the inner engine pulsing/flaring activity. The lack of detection could suggest that this mechanism is not occurring, or alterations to our assumptions are required. One of the model dependencies is the fraction of energy in the magnetic field, and following the assumptions of the magnetic wind model this is assumed to be $\epsilon_{B} = 10^{-3}$ \citep{Katz1997}. If we instead make a more conservative estimate of $\epsilon_{B} \lesssim 10^{-4}$ we would predict radio emission to peak in brightness at $\sim$19\,mJy, approximately the rms noise in the images, and thus would not expect to see the radio pulses in this case. This suggests that our non-detection may favour a lower fraction of energy in the magnetic field for this GRB. In the case of rapid radio follow-up of long GRB 210419A, the non-detection of emission related to X-ray flares results in an upper limit corresponding to a constrain of $\epsilon_{B} \lesssim 10^{-3}$ \citep{Tian2022b}. In a recent search for coherent radio emission from SGRBs, \citet{Tian2022a} suggest constraints of $\epsilon_{B} \lesssim 2 \times 10^{-4}$ in the GRB jet from similar arguments. A wide range of values between 10$^{-5}$ and 0.33 have been estimated for pre-\swift\ GRBs predominantly through afterglow light curve modelling \citep{Panaitescu2001,Santana2014}, and this has been extended down to $10^{-8}$ in a \swift\ study \citep{Santana2014}. It is one of the hardest parameters to pin down using afterglow modelling, using different methods on the same dataset can yield quite different results.
    
    Certainly, the greatest uncertainty lies in the redshift estimate. In this case the lack of a spectroscopic redshift leads us to adopt a redshift value of $z=2$ from multi-colour afterglow photometry \citep{gcnKann2021a}, which happens to lie at approximately the mean redshift for \swift\ long GRBs \citep{Evans2009}. This has implications on the observed brightness and, importantly, the duration and timing of the flares. We calculated the smallest redshift that would allow us to still observe the peak of the first flare, and this is $z = 1.914$, smaller values would cut off part of the predicted pulse and the expected brightness would drop below the noise threshold. The similar limit for the third flare is $z=1.637$, but we show this flare would not be observable regardless, given the noise in the image. The upper limit of redshift on our observations extends far beyond our assumed value: the LOFAR window extends to cover a much broader range of dispersion delays, but at $z\sim6$, we expect the radio flux to drop below our sensitivity thresholds. We therefore examined the host galaxy through published photometry and additional observations with the DOT telescope (Section \ref{sec:optical_observations}). We used the \textsc{Code Investigating GALaxy Emission} \citep[\textsc{CIGALE},][]{Boquien2019} package to evaluate the galaxy's spectral energy distribution (SED). \textsc{CIGALE} uses a series of additive templates to simulate the full emission of a galaxy and here we use a double exponential star formation history and a low metallicity stellar population \citep{Bruzual2003} to model the starburst galaxies typical of LGRB hosts. Dust attenuation was included using the \citet{Calzetti2000} extinction law and \textsc{CIGALE} assumes the cosmology identified by the Seven-year \textit{Wilkinson Microwave Anisotropy Probe} observations \citep[$H_{\textrm{0}} = 70.4$ km s$^{-1}$ Mpc$^{-1}$ and $\Omega_{\textrm{M}} = 0.272$; ][]{Komatsu2011}. \textsc{CIGALE} creates a grid of models across the combined parameter space of the component templates and uses Bayesian inference to derive the most likely values for each parameter. We initially allowed the redshift to vary between 0.5 and 2.5, finding it tended towards a lower redshift of $z\sim1.3$ but this is poorly constrained and likely to be unreliable. We therefore assumed a fixed redshift of $z=2$ from \citet{gcnKann2021a} and found it to be compatible with the host data. We infer a stellar mass of $\log(M_*/M_{\odot}) \sim 10.9$, a star formation rate of $\sim40$ $M_{\odot}$ yr$^{-1}$ and a dust extinction of $E(B-V)\sim0.1$, properties generally consistent with the known LGRB host population \citep[e.g.][]{Levesque2014,Stanway2014}. We note, however, that these values are generally not tightly constrained due to the narrow wavelength range of our host galaxy photometry. This star formation rate can also be used to derive an estimate of the expected 150 MHz flux density. Following the prescription of \citet{Gurkan2018} we find a value of $\sim0.015$ mJy, which aligns well with our non-detection of persistent emission in the 2-hour HBA observation in which the limiting flux density is 3 mJy.

    Currently LOFAR is limited to the use either 48 HBA or 48 equivalent Low Band Antenna (LBA) at a time, and thus our rapid response observations only make use of the HBA. In this case we are further from the peak radio frequency, but the sensitivity is considerably better such that it is worth the trade-off. The upgrades to LOFAR 2.0 in the future will allow 96 antennas with simultaneous observations using both bands (Hessels et al., in prep). The LBA operates at 10--90\,MHz meaning we can probe closer to the predicted radio flux peak at $\sim 1$\,MHz. Given this, in the case of GRB 210112A, we should expect to observe at least the brightest flare even with the sensitivity of LBA not being quite as deep as HBA. The rapid response constraints also loosen for a lower observing frequency, where the same $z=2$ dispersion delay applied to LBA is 3415\,s, placing it in the middle of our LOFAR observing window. A detection in both frequencies for a given equivalent X-ray flare may allow us to begin to constrain the $\beta$ parameter, the high frequency tail power law index.

    Going forward, it's clear that with uncertain model parameters, uncertain redshift, and the unpredictability of X-ray flares occurring during the LOFAR observation, if any occur at all for a given GRB, that the best strategy is to observe a large sample of bursts in this way. Sample parameter distributions can therefore be folded through the observed flux limit distributions, to give proper statistics.

\section{Conclusion}
\label{sec:concl}

    In this paper, we present the observations of GRB\,210112A made with LOFAR's HBA in rapid response mode. We test a magnetic wind model, where we look for predicted low-frequency coherent radio emission associated with the prompt phase $\gamma$-ray pulses and X-ray flares. A detection at the predicted time would be evidence the GRB jet is Poynting flux dominated. Using LOFAR HBA observations, we expect to at least see the first flare given the model and redshift assumptions.

    However, no emission is detected at the position of the GRB on short imaging timescales. We also probe the full 2 hour observation for any persistent emission and produce a 3\,mJy upper limit, which is consistent with the expected emission from a typical LGRB host galaxy at a redshift of 2.

    A non-detection in one or a few sources does not necessarily rule out this magnetic wind model. We have discussed uncertainties in our assumptions and the model itself. The redshift is obtained tentatively, through multi-colour afterglow photometry, though it is consistent with spectral fitting and host galaxy template fitting. The effect of redshift is primarily on the predicted duration and timing of the radio flares. At $z<1.91$ the brightest flare would fall before our observation window. Our search is optimised for $z=2$, but if this differs then it has implications on our sensitivity to a dispersed signal of a different length. Additionally, the emission models and input parameters still remain uncertain. Changes in the strength of magnetic field or efficiency of the emission mechanisms means less radio emission is expected, possibly low enough to not be detected in our images. At $\epsilon_{B} = 10^{-4}$, the predicted radio flux would be below the image noise in our observations.

    To confidently test the prediction of radio emission from a magnetic wind model, we need several rapid follow-up observations of GRBs, simultaneous with \swift\ or similar X-ray prompt follow-up, and with spectroscopic redshifts. Given our assumed DM = $1200z$, $z$ is directly proportional to the dispersion delay of radio pulses. Hence, a response time of 5 minutes after the initial \swift-BAT trigger allows us to probe radio emissions to bursts at $z > 1$. This hard limit is relaxed slightly further as we probe the more accessible X-ray flares, which usually occur in the first few hundred seconds of the GRB, during or after the prompt emission. The Space Variable Objects Monitor \citep[SVOM,][]{Atteia2022} mission is expected to launch mid 2023. The ECLAIRs \citep{Godet2014} telescope onboard will provide GRB triggers at a low energy threshold of 4\,keV, lower than that of existing missions. This will provide access to the soft X-ray regime of prompt emissions, cementing the links between X-ray flare and $\gamma$-ray pulses. In particular, with LOFAR 2.0 rapid observations, we will be able to probe multiple radio frequency channels simultaneously.

\section*{Acknowledgements}
We thank the referee Jonathan Katz for critically reviewing this work. The LOFAR data were taken under programme LC15$\_$013 (PI Starling). AH is supported by an STFC studentship. RLCS acknowledges support from an ASTRON Helena Kluyver visiting fellowship. AR acknowledges funding from the NWO Aspasia grant (number: 015.016.033). IdR acknowledges support through the project CORTEX (NWA.1160.18.316) of the research programme NWA-ORC which is (partly) financed by the Dutch Research Council (NWO). RAJEF acknowledges support from the UK Space Agency and the European Union’s Horizon 2020 Programme under the AHEAD2020 project (grant agreement number 871158). KW acknowledges support through a UK Research and Innovation Future Leaders Fellowship awarded to Dr.~B.~Simmons (MR/T044136/1). 
This work made use of data supplied by the UK \textit{Swift} Science Data Centre at the University of Leicester. We thank the staff of the 3.6m Devasthal Optical Telescope (DOT), an optical telescope run and managed by Aryabhatta Research Institute of Observational Sciences (ARIES), an autonomous Institute under the Department of Science and Technology, Government of India, for service observations with the ADFOSC instrument. 
This paper is based (in part) on data obtained with the International LOFAR Telescope (ILT) under project code LC10 012. LOFAR is the Low Frequency Array designed and constructed by ASTRON. It has observing, data processing, and data storage facilities in several countries, that are owned by various parties (each with their own funding sources), and that are collectively operated by the ILT foundation under a joint scientific policy. The ILT resources have benefited from the following recent major funding sources: CNRS-INSU, Observatoire de Paris and Universit\'{e} d'Orl\'{e}ans, France; BMBF, MIWF-NRW, MPG, Germany; Science Foundation Ireland (SFI), Department of Business, Enterprise and Innovation (DBEI), Ireland; NWO, The Netherlands; The Science and Technology Facilities Council, UK; Ministry of Science and Higher Education, Poland.

%%%%%%%%%%%%%%%%%%%%%%%%%%%%%%%%%%%%%%%%%%%%%%%%%%
\section*{Data Availability}

LOFAR data are available in the LOFAR Long-Term Archive (LTA) at \url{https://lta.lofar.eu}.
Swift data and XRT products are available via the UK Swift Science Data Centre (UKSSDC) at \url{www.swift.ac.uk}.

%%%%%%%%%%%%%%%%%%%% REFERENCES %%%%%%%%%%%%%%%%%%

\bibliographystyle{mnras}
\bibliography{references,ref_gcn}

%%%%%%%%%%%%%%%%% APPENDICES %%%%%%%%%%%%%%%%%%%%%

\appendix

\newpage
\section{Spectral Fits}

\renewcommand{\arraystretch}{1.5}
\begin{table*}
\caption{Time coverage and spectral fits for all timesliced phases of BAT and XRT data. Where there is simultaneous coverage, fits are performed with a broken power law covering both sets of data. In these cases, $\Gamma_{1}$ the lower energy photon index and $\Gamma_{2}$ the higher energy photon index. Errors shown are 90\% confidence intervals. BAT only phases are fitted with a cut-off power law. XRT only phases are fitted with a double absorbed power law. \textbf{Phase 4} is the spectral fit used to constrain and freeze the intrinsic absorption parameter to $2.980 \times 10^{22}$ cm$^{-2}$.}
\label{tab:spectralfits}
\begin{tabular}{llrrccc}
\hline
Phase         & Instrument(s) & Start (sec) & Stop (sec) & $\Gamma_{1}$           & $\Gamma_{2}$           & Chi-Square / Degrees of Freedom \\ \hline
Preflare      & BAT & 0        & 37.72    & -                      & $0.67^{+0.36}_{-0.41}$ & 63.18 / 55 \\
Flare 1       & BAT & 37.72    & 89.94    & -                      & $1.20^{+0.02}_{-0.03}$ & 33.95 / 55 \\
Flare 1 Rise  & BAT & 37.72    & 65.48    & -                      & $1.16^{+0.03}_{-0.07}$ & 48.89 / 55 \\
Flare 1 Peak  & BAT & 65.48    & 75.91    & -                      & $1.15^{+0.02}_{-0.06}$ & 23.94 / 55 \\
Flare 1 Decay & BAT, XRT      & 78.35       & 89.94      & $0.83^{+0.06}_{-0.06}$ & $1.74^{+0.05}_{-0.05}$ & 450.67 / 607              \\
Flare 2       & BAT, XRT      & 89.61       & 117.8      & $0.85^{+0.03}_{-0.03}$ & $1.72^{+0.04}_{-0.04}$ & 727.31 / 792              \\
Flare 3       & BAT, XRT      & 117.8       & 153        & $0.48^{+0.13}_{-0.16}$ & $1.54^{+0.02}_{-0.02}$ & 679.08 / 653              \\
Phase 2       & BAT, XRT      & 153         & 614        & $1.77^{+0.04}_{-0.04}$ & $2.22^{+0.06}_{-0.05}$ & 743.80 / 774              \\
Phase 3 WT    & XRT & 614      & 935      & $2.34^{+1.16}_{-1.14}$ & -                      & 0.59 / 3 \\
Phase 3 PC    & XRT & 935      & 8913     & $1.96^{+0.05}_{-0.05}$ & -                      & 273.47 / 365 \\
\textbf{Phase 4}       & \textbf{XRT} & \textbf{8913}     & \textbf{$\mathbf{1.70\times10^{5}}$} & \textbf{$\mathbf{2.15^{+0.05}_{-0.05}}$} & \textbf{-}                     & \textbf{419.08 / 413} \\
Phase 5       & XRT & $1.70\times10^{5}$ & $1.08\times10^{6}$ & $1.74^{+0.28}_{-0.26}$ & -                      & 405.04 / 429 \\ \hline
\end{tabular}
\end{table*}
\renewcommand{\arraystretch}{1}

\section{Temporal Fits}

\renewcommand{\arraystretch}{1.5}
\begin{table*}
\caption{Temporal fits corresponding to \autoref{fig:laff_fit}. Phase 0 is poorly fit as there are few data points remaining for continuum fitting after flaring data has been excluded.}
\label{tab:temporalfits}
\begin{tabular}{lrr}
\hline
Canonical Phase    & Temporal Index $\alpha$          & Post-Phase Break Time (s) \\ \hline
Phase 0            & $0.49 \pm 0.95$   & $83.68 \pm 318.6$         \\
Phase 1            & $1.91 \pm 0.03$   & $604.8 \pm 23.2$          \\
Phase 2            & $0.73 \pm 0.03$   & $7333 \pm 1237$           \\
Phase 3            & $1.30 \pm 0.05$   & $157574 \pm 67602$        \\
Phase 4            & $1.86 \pm 0.34$   & -                         \\ \hline
\multicolumn{2}{l}{Normalisation}      & 1.14 $(\pm0.04) \times 10^{-7}$ \\
\multicolumn{2}{l}{Chi-square}         & 318.6                     \\
\multicolumn{2}{l}{Reduced Chi-square} & 1.3                       \\ \hline
\end{tabular}
\end{table*}
\renewcommand{\arraystretch}{1}

\section{Optical Data}

\renewcommand{\arraystretch}{1.5}
\begin{table*}
\caption{Table showing the optical data plotted in \autoref{fig:joint_lc} from several observatories.}
\label{tab:opticaldata}
\begin{tabular}{lccrrl}
\hline
Telescope & Filter & Time since BAT Trigger (days) & Magnitude (AB)     & Flux ($\mu$Jy)                           & Reference             \\ \hline
OSN       & Rc     & 0.126                         & 22.41 $\pm$ 0.12   & 3.44 ($\pm$ 0.36)   & \citet{gcnKann2021a}  \\
OSN       & Ic     & 0.126                         & 21.10 $\pm$ 0.075  & 9.44 ($\pm$ 0.63)    & \citet{gcnKann2021b}  \\
Mondy     & Rc     & 0.807                         & $>22.40$           & $<3.38$               & \citet{gcnPankov2021} \\
DFOT      & Rc     & 0.937                         & $>22.69$           & $<2.59$               & \citet{gcnGupta2021}  \\
OSN  & Ic & 1.13   & 23.02 $\pm$ 0.217 & 1.61 ($\pm$ 0.29)  & \citet{gcnKann2021b}  \\
DOT  & i' & 1.93  & 23.19 $\pm$ 0.166 & 1.98 ($\pm$ 0.28)   & \citet{gcnDimple2021} \\
CAHA & Ic & 5.17 & 23.52 $\pm$ 0.19  & 1.02 ($\pm$ 0.16)  & \citet{gcnRossi2021}  \\
LBT       & z'     & 24.3                          & 23.38 $\pm$ 0.15   & 1.62 ($\pm$ 0.21)  & \citet{gcnRossi2021}  \\
LBT       & r'     & 24.3                          & 23.96 $\pm$ 0.15   & 0.98 ($\pm$ 0.13)   & \citet{gcnRossi2021}  \\
DOT       & r'     & 771.0                         & 23.89 $\pm$ 0.117 & 1.05 ($\pm$ 0.11)  & \it{this work}              \\
DOT       & i'     & 771.0                         & 23.67 $\pm$ 0.231 & 1.27 ($\pm$ 0.25)  & \it{this work}              \\
DOT       & z'     & 778.0                         & 23.35 $\pm$ 0.231 & 1.66 ($\pm$ 0.34)  & \it{this work}              \\ \hline
\end{tabular}
\end{table*}
\renewcommand{\arraystretch}{1}

% Don't change these lines
\bsp	% typesetting comment
\label{lastpage}
\end{document}